\documentstyle[prd,aps,preprint]{revtex}

\begin{document}
\title{Chaos in Non-Abelian Gauge Fields, Gravity, and Cosmology
 \thanks{Talk presented at the Ninth Marcel Grossmann Conference
 (MG9), Rome, Italy, July 2000}}
\author{S.G. Matinyan\thanks{Present address: 3106 Hornbuckle Place,
 Durham, NC 27707.}\\
 Department of Physics, Duke University, Durham, NC 27708-0305}
\maketitle

\begin{abstract}
This talk describes the evolution of studies of chaos in Yang-Mills 
fields, gravity, and cosmology. The main subject is a BKL regime near
the singularity $t=0$ and its survival in higher dimensions and in
string theory. We also describe the recent progress in the search for
particle-like solutions of the Einstein-Yang-Mills system (monopoles
and dyons), colored black holes and the problem of their stability.
\end{abstract}

\section{Introduction}

The great achievements of theoretical physics in the 20th century,
Einstein's theory of general relativity and the Yang-Mills theory
of non-Abelian gauge fields, have many common properties: both are
gauge theories and they are, in contrast to Maxwell's theory of the
electromagnetic field, nonlinear. From this point of view, the 
chaoticity of the corresponding fields is not a surprise. On the
other hand, there are many examples of stable solutions (solitons)
of nonlinear field equations. Thus, the question of chaos in the
equations of general relativity (GR) and the Yang-Mills (YM) equations
is not a trivial and straightforward one. The evolution of our insight
into chaos in the YM theory provides a good illustration for this
nontriviality.

In the 1970s the problem of the integrability of the classical
YM equations was actively studied. All attempts to find additional
(to the trivial ones, such as energy) integrals of the YM equations
in Minkowski space proved unsuccessful. This gave rise to a progran
aimed at describing the YM dynamics not in terms of the potentials
and fields, but rather in terms of the loop (string) variables
\cite{ref1}. In a Euclidean space, the famous classical solutions
called instantons \cite{ref2}, were already known. But what was
going on in (3+1)-dimensional space-time? Why are soliton-like
solutions absent there? The answer is quite simple: The YM equations
are chaotic even, as it is shown below, in the simplest conceivable
limiting cases.

\section{Chaos in the Classsical YM System}

Consider a spatially homogeneous, sourceless YM field with
potentials $A_i^a(t) (i,a=1,2,3)$ in the $A_0^a=0$ gauge, for
the gauge group SU(2). Such vector potentials can be considered
as the relevant degrees of freedom in the infrared limit. They
also may describe field configurations with a high gluon density
(see ref. \cite{ref3} for details). With these approximations,
the YM equatins are reduced to a discrete Hamiltonian dynamical
system:
\begin{equation}
{d^2A_i^a \over dt^2} 
- g^2 (A_j^a A_j^b A_i^b - A_j^b A_j^b A_i^a) = 0
\label{eq1}
\end{equation}
with the conserved external and internal angular momenta:
\begin{eqnarray}
M_i = \epsilon_{ijk} A_j^a {\dot A}_k^a , \nonumber\\
N^a = \epsilon^{abc} A_i^b {\dot A}_i^c .
\label{eq2}
\end{eqnarray}
These quantities vanish for the sourceless field. Thus there exist,
including the energy, seven integrals of motion, but the system is
still not integrable. In the simplest nontrivial case, with
$x=gA_1^1$, and $y=gA_2^2$ as the nonvanishing components, we obtain
a system of two nonlinearly coupled oscillators with the Hamiltonian
(the socalled $x^2y^2$ model \cite{ref4}):
\begin{equation}
H = {1\over 2} (p_x^2 + p_y^2 + x^2y^2)
\label{eq3}
\end{equation}
which exhibits strong chaotic behavior: most periodic orbits are
unstable, one Lyapunov exponent is positive, etc. (see ref.
\cite{ref3} for details).

There are several mechanisms that can eliminate this chaoticity
\cite{ref3}. One that is especially important here, is the coupling
to a scalar field (Higgs field) with nonvanishing vacuum expectation
value $\langle\phi\rangle = v$. Then a term proportional to
$g^2v^2(x^2+y^2)$ is added to the Hamiltonian (\ref{eq3}), which
reduces the system to one of two weakly coupled harmonic oscillators
when the vacuum expectation value $v$ is large. However, at small
enough $v$ the chaos persists, and the transition between chaotic
and regular behavior is controlled by the parameter $g^2v^2/E$
where $E$ is the energy density \cite{ref5}.

The chaoticity of the sourceless YM field theory was also 
established for the special case of spherically symmetric
field configurations \cite{ref6}. The systematic stuy of the
sourcesless classical lattice gauge theory by the Duke group
also found evidence for spatio-temporal chaos in this system
(see ref. \cite{ref3} for details).

Concluding this brief survey of the study of chaos in the YM
fields, I need to emphasize that these results demonstrate that
the classical YM fields lack any special stable configurations.
All states are chaotic and no particular configuration dominates,
in contrast to the Euclidean case, where the instantons make a
dominant contribution to the functional integral. It is also
noteworthy that the YM field is a dynamical system where quantum
corrections are undoubtedly important. These corrections will
lead to the elimination of most, but not all, properties that
are characteristic of the chaotic YM system. This holds true
for the YM quantum mechanics as well as for the quantum YM
field theory \cite{ref7,ref8}.

\section{Chaos in Gravity and Cosmology}

\subsection{BKL scenario of the approach to the singularity}

The study of chaos in general relativity (GR) has an even
longer history than in the case of the YM field. Einstein himself
believed that, due to the highly nonlinear self-interaction of the
gravitational field, the equations of motion of GR cannot be solved
in general. This is the reason why he was so pleased when Karl
Schwarzschild found the general solution with spherically symmetry.
In fact, due to its different gauge group structure, gravity has 
more "capacity to resist" chaos than the YM field. From this point
of view, and to exhibit the closest similarity to the spatially
homogeneous YM field, let us consider the Einstein equations near 
the space-time singularity in the synchronous coordinate frame,
where the metric $g_{\mu\nu}$ can be expressed as a function of
the synchronous time $t$ only.

Landau remarked (see ref. \cite{ref9}) that 
$g\equiv \det(g_{\mu\nu})$ tends to zero as $t\to 0$, at some
{\em finite} time interval, independent of the equation of state
or the character of the gravitational field. The behavior of the
metric $ds^2 = dt^2 - d\ell^2$ with ($\alpha,\beta=1,2,3$)
$d\ell^2 = \gamma_{\alpha\beta} dx^\alpha dx^\beta$ near the
singularity at $t=0$ has the form
\begin{equation}
d\ell^2 = t^{2p_1} dx_1^2 + t^{2p_2} dx_2^2 + t^{2p_3} dx_3^2 
\label{eq4}
\end{equation}
with $p_i$ being the eigenvalues of the matrix 
$\lambda_\alpha^\beta$ in the equation of motion:
$$
{\dot\gamma}_{\alpha\beta} = {2\over t} \lambda_\alpha^\delta
\gamma_{\delta\beta} .
$$
Since $\lambda$ satisfies the trace relations
${\rm tr}(\lambda) = {\rm tr}(\lambda^2) = 1$, the exponents obey
the conditions
$\sum_{i=1}^3 p_i = \sum_{i=1}^3 p_i^2 = 1$.
Equation (\ref{eq4}) is the so-called Kasner solution \cite{ref10}
corresponding to the Bianchi I geometry.
Thus, in this simplest case, we have the regular flat, homogeneous,
but anisotropic space with the total volume homogeneously
approaching the singularity ($dx_1dx_2dx_3 \sim t$).

A generally chaotic oscillatory behavior was observed for the
Bianchi IX geometry in the pioneering study of Belinski,
Khalatnikov, and Lifshitz \cite{ref9,ref11} of the Mixmaster
universe \cite{ref12}. This study greatly stimulated the 
interest in the problem of chaos in GR. For the Bianchi IX
geometry, a generalization of (\ref{eq4}):
\begin{equation}
d\ell^2 = (a^2 \ell_\alpha \ell_\beta + b^2 m_\alpha m_\beta
+ c^2 n_\alpha n\beta) dx^\alpha dx^\beta
\label{eq5}
\end{equation}
where $\vec\ell$, $\vec m$, $\vec n$ are unit vectors along the
principal axes, the coefficients $a,b,c$ are functions of $t$.
The evolution towards the singularity at $t=0$ proceeds via
a series of successive oscillations, during which the distances
along two of the principal axes oscillate, while they shrink
monotonically along the third axis (Kasner epochs). The volume
$V=16\pi^2 abc$ again drops approximately as $t$. A new ``era''
begins when the monotonically falling metric component begins
to oscillate, while one of the previously oscillating directions
begins to contract. This approach to the singularity, revealing
itself as an infinite succession of alternating Kasner epochs,
asymptotically takes on the character of a random process.

In the study of the Mixmaster universe it is useful to employ
its Hamiltonian formulation \cite{ref12}, defined in the
mini-superspace of the variables $(\Omega, \beta_+, \beta_-)$
where $\Omega$ is the volume and the $\beta_\pm$ are connected
to the coefficients $a^2,b^2,c^2$ as follows:
\begin{equation}
\gamma_{\alpha\beta} = {\rm diag}(a^2,b^2,c^2) =
e^{2\Omega} \left(e^{2\beta}\right)_{\alpha\beta}.
\label{eq6}
\end{equation}
Obviously, ${\rm tr}(\beta)=0$. In this approach, the spatial
scalar curvature $V(\beta_+,\beta_-)$ plays and essential role
as the two-dimensional time-dependent potential in the
Hamiltonian
\begin{equation}
2H = -p_\Omega^2 + p_+^2 + p_-^2 + e^{-4\Omega}
\left(V(\beta_+,\beta_-) -1\right)
\label{eq7}
\end{equation}
with $p_{\Omega,+,-}$ being the momenta conjugate to the
variables $\Omega,\beta_+,\beta_-$. The time coordinate $\tau$
is defined by $dt = abc d\tau$, where $t$ is the synchronous time.
For the Bianchi geometry, $V(\beta_+,\beta_-)=0$.

The evolution governed by epochs and eras here corresponds
to bounces off the potential $V$ which has three corners in the
plane $(\beta_+,\beta_-)$ for fixed $\Omega$. The epochs
correspond to bounces outward along the corner of the potential,
while a new era begins with the change of a corner. Qualitatively,
the picture resembles the behavior found in the $x^2y^2$ model
of YM dynamics, where the motion along the channels formed by
the equipotential lines (hyperbolae) corresponds to an ``epoch'',
and the change of the channel starts a new ``era'' (see ref.
\cite{ref3} for details of the YM case).

Concluding this subsection, we remark that a powerful tool for
the study of chaos, the Lyapunov exponents $\lambda$, sometimes
does not apply in the case of gravity, leading to a contradiction
between the analytical statements of the BKL type and numerical
calculations. This is not surprising since $\lambda$, measuring
the rate of divergence between neighboring trajectories, depends 
on the choice of the time variable. E.g., $\lambda=0$ in the BKL
time variable $\tau = \int dt/abc$, while $\lambda=\infty$ in
the metric time $t$ \cite{ref13,ref14}. One needs to have a
covariant measure of chaos that does not depend on the choice 
of the time variable. For instance, the notion of integrability,
i.e. the existence of first integrals of motion, is appropriate. 
Concerning the cases discussed above, it has been shown that
the Bianchi type VIII and IX models do not possess such integrals
\cite{ref15}.

\subsection{BKL problem with a scalar field}

It was shown \cite{ref16} in 1973 (see also the recent paper
\cite{ref17}) that the addition of a massless scalar field
eliminates chaos in the BKL problem. For illustration, consider
the simpler case a scalar field conformally coupled to gravity
in the Friedman-Robertson-Walker (FRW) universe. In the
Hamiltonian version we have after proper rescaling of the
scalar field $\phi$:
\begin{equation}
H(a,\phi;p_a,p_\phi) = {1\over 2}[-p_a^2 - k a^2
+p_\phi^2 + k \phi^2 + m^2\phi^2a^2]
\label{eq8}
\end{equation}
with $a(t)$ the FRW scale parameter and $k=\pm 1,0$. (Compare
(\ref{eq8}) with the Hamiltonian of the $x^2y^2$ model of the
YM-Higgs system \cite{ref3}. The strange sign of the variables
associated with the FRW scale factor $a$ is specific to gravity.)
Obviously, chaos does not appear for the massless scalar field
($m=0$). For $m\neq 0$, one finds, in general, chaotic behavior
due to the coupling between the ``oscillators'' $a(t)$ and
$\phi(t)$.

If one adds to (\ref{eq8}) the self-interaction of the scalar 
field $\phi$ and the contribution from the cosmological constant
${1\over 4}(\lambda\phi^4+\Lambda_0a^4)$ (where again $\phi$ is
the rescaled scalar field and $\Lambda_0$ is the rescaled
cosmological constant \cite{ref18}), then the Painlev\'e analysis
of the equation of motion
\begin{eqnarray}
{\ddot a} + ka - m^2a\phi^2 - \Lambda_0a^3 = 0 \nonumber\\
{\ddot\phi} + k\phi + m^2a^2\phi + \lambda\phi^3 = 0
\label{eq9}
\end{eqnarray}
leads to the conclusion that the integrability test is passed 
only for a few special values of $\Lambda_0$, $\lambda$, $m^2$
and $k$ \cite{ref18} (see also ref. \cite{ref19,ref20}).
The case $m=0$ in (\ref{eq9}) gives the reguar regime:
\begin{eqnarray}
a(t) = \sqrt{2k\over -\Lambda_0} (\cosh\sqrt{k}t)^{-1}
\nonumber\\
\phi(t) = \sqrt{2k\over \lambda} (\cosh\sqrt{k}t)^{-1}.
\label{eq10}
\end{eqnarray}

\subsection{Higher dimensions}

{From} the BKL scenario we have seen that, as one approaches the
singularity at $t=0$, spatial dimensions ($d=3$) are continuously
mixed by the dynamics in an infinite series of oscillations. This
mixing, as well as the successive length of the oscillatory periods
(i.e. the number of Kasner epochs in each era), have the character 
of a random process \cite{ref11}. All spatial dimensions stand on
an equal footing. But what happens if one increases, in the spirit
of the modern development of Kaluza-Klein models and string theory
motivated models, the number of spatial dimensions?

The problem can have importance in the context of the dimensional
reduction that occurs with compactification. Consider the Kasner
solution which generalizes the metric (\ref{eq4}) for a space-time
of dimension $D=d+1$:
\begin{equation}
ds^2 = dt^2 - \sum_{i=1}^d t^{2p_i(x)} (\omega_i)^2 .
\label{eq11}
\end{equation}
with $p_i(x)$ time independent and satisfying Kasner's conditions
\begin{equation}
\sum_{i=1}^d p_i(x) = \sum_{i=1}^d p_i(x)^2 = 1 ,
\label{eq11a}
\end{equation}
and time-independent one-forms $\omega_i(x) = \ell_j^i(x) dx^j$
with $\det(\ell_j^i) \neq 0$.

Similar arguments as for $d=3$ show that for $d\geq 10$ the monotonic
approach to $t=0$ according to (\ref{eq11}) has the same degree of generality 
as the BKL oscillatory regime. They are based on the possible neglect
of the spatial gradients (present in the spatial curvature) compared
with the terms involving time derivatives, which are always of the
relative magnitude $t^{-2}$ in the vacuum Einstein equations, for any
value of $d$. Thus, one requires:
\begin{equation}
\lim_{t\to 0} {t^2} ~^{(d)}R_k^i = 0 ,
\label{eq12}
\end{equation}
where $^{(d)}R_k^i$ is the $d$-dimensional spatial Ricci tensor.
This leads to the condition for the ``Kasner stability region'':
\begin{equation}
\alpha_{ijk} \equiv 2p_i + \sum_{\ell\neq i,j,k} p_\ell > 0
\label{eq13}
\end{equation}
for $i,j,k$ all different. When (\ref{eq13}) contradicts the
Kasner conditions (\ref{eq11a}), we unavoidably have the BKL
scenario. As it was shown in ref. \cite{ref21}, this indeed
occurs for $d<10$. However, for $d\geq 10$, the conditions
(\ref{eq13}) are not contradictory and define an open region
of the so-called ``Kasner sphere'' (\ref{eq11a}).

Thus the generalized Kasner solution (\ref{eq11}) with monotonic
power-law behavior of the spatial distances, becomes a general
solution of the Einstein equations near a cosmological singularity
when $D=d+1\geq 11$. Moreover, in this case the chaotic regime
of BKL type becomes unstable and is replaced by the monotonic
regime (\ref{eq11}) \cite{ref22,ref23,ref24}. Whereas for $d<10$
all spatial dimensions appear on an equal footing, as remarked
earlier, the spatial directions are split into two groups for
$d\geq 10$. One group is expanding, the other one contracting.
Thus, some dimensions dominate over others as the singularity is
approached, which may provide a ``natural'' explanation for the
compactification to a lower number of dimensions near the
singularity.

\subsection{Superstring theory and the BKL regime}

Superstring theories with their rich field content (graviton,
dilaton, other scalars, vectors, $p$-forms, etc.) and large
number of dimensions ($D=10,11$) may modify the BKL regime
even at the classical level \cite{ref25}. In particular, the
dilaton affects the formation of the Kasner sphere. Instead
of (\ref{eq11a}), we have:
\begin{equation}
p_\phi^2 + \sum_{i=1}^d p_i^2 - \left( \sum_{i=1}^d p_i \right)^2
=0 ; \qquad \sum_{i=1}^d p_i = 1 .
\label{eq14}
\end{equation}
Two additional sets of stability constraints, called electric
and magnetic components and related to the contribution of the 
$p$-form fields $A_p$ to the Einstein-dilaton equations appear, 
in addition to the pure gravitational stability conditions 
(\ref{eq13}). The positivity of these constraints, together
with (\ref{eq13}) ensures the stability of the generalized
Kasner solution. The result of the study \cite{ref25} is that
for all superstring models there exists no open region of the
Kasner sphere (\ref{eq14}) where {\em all} exponents are positive.
In other words, the generic solution near $t=0$ for massless
bosonic degrees of freedom exhibits the BKL behavior.

Three ingredients were used to obtain this result: ({\em i})
inclusion of the $p$-forms in the field content of the theory;
({\em ii}) a sufficiently strong coupling of the dilaton to
the graviton; and ({\em iii}) string dualities.\footnote{We
note that there exists an opposite statement in the literature
\cite{ref26} that the duality symmetry appears to be incompatible
with the presence of the chaotic behavior near $t\to 0$. This
statement is based on the qualitative Hamiltonian description
of the Mixmaster models.} The limitation of this study is the
restriction of its arguments to the tree level.

In conclusion, we remark that if the BKL-type regime survives
in string theories, then the spatial inhomogeneity continuously
increases toward a singularity \cite{ref25} (``turbulent''
universe \cite{ref27,ref28}).

\section{Solitons and Black Holes in the Einstein-YM System}

Recently, the interplay of gravity and the YM system was studied
very actively. It is known that the sourceless YM equations have 
no particle-like solutions in Minkowski space \cite{ref29,ref30}.
The chaotic properties of YM dynamics ensure that no particular
configuration dominates. In the pioneering paper \cite{ref31}
numerical evidence was presented for the existence of a discrete 
family of globally regular solutions of the static Einstein-YM 
equations for the YM gauge group SU(2)).

The existence of particle-like solutions of the Einstein-YM system
is a result of the balance between the YM repulsive force and the
gravitational attraction. The situation is similar to the addition
of a scalar field to the free YM field resulting in the formation
of the 't Hooft -- Polyakov monopole solution. Complementary to
the study \cite{ref31}, the existence of black hole solutions
of the EYM system was established in ref.~\cite{ref32} (colored
black holes). This result contradicts the notion of the ``baldness''
of black holes (see e.g. \cite{ref33}) and hints at the instability
of the colored black hole solutions.

Indeed, it was shown by various authors 
\cite{ref34,ref35,ref36,ref37,ref38,ref39,ref40,ref41,ref42} that 
the particle-like solutions of the type \cite{ref31}, as well as the
non-Abelian SU(2) EYM black holes \cite{ref32}, are unstable for
$\Lambda \geq 0$ (flat space or de-Sitter space). Surprisingly, 
the situation is different for the anti-de-Sitter case, $\Lambda<0$.
It was shown that a continuous class of regular monopole and dyon
\cite{ref43} solutions without cosmological horizon and ``hairy''
black hole solutions \cite{ref44} exist in the EYM theory in any
space that is asymptotically anti-de-Sitter (adS). For solutions
without nodes in the non-Abelian magnetic field, the stability
was established. The stability of nodeless dyon solutions is
awaiting further study.

The fact that for $\Lambda<0$ there exists a continuous set of
such solutions, whereas for $\Lambda\geq 0$ the set is discrete,
invites the conjecture \cite{ref43} that the moduli space of
solutions has a fractal structure. In conclusion, the existence
of stable particle-like configurations may have important
consequences for the evolution of the early universe if it ever
passed through a adS phase.

\section*{Acknowledgments}

It is my great pleasure to thank Remo Ruffini and Vahagn
Gurzadyan for the invitation to the superbly organized MG-9
conference and the support which made my participation possible.
I thank Berndt M\"uller for numerous stimulating discussions
on the chaos in particle physics and general relativity.

\end{document}